\documentclass[twocolumn]{aastex61}

\usepackage{savesym}
\savesymbol{tablenum}
\usepackage{siunitx}
\restoresymbol{SIX}{tablenum}
\usepackage{amsmath}

\newcommand{\oiii}{[O III]}
\newcommand{\oiiifull}{\oiii$\lambda$5007\AA}
\renewcommand{\d}[2]{\frac{d#1}{d#2}}

\shorttitle{The NLR Size-Luminosity Relationship}

\begin{document}

\title{The Size-Luminosity Relationship of Quasar Narrow-Line Regions}

\author{Ross Dempsey}
\affiliation{Department of Physics and Astronomy, Johns Hopkins University, 3400 N. Charles St., Baltimore, MD 21218, USA}
\author{Nadia L. Zakamska}
\affiliation{Department of Physics and Astronomy, Johns Hopkins University, 3400 N. Charles St., Baltimore, MD 21218, USA}



\begin{abstract}
The presence of an active galactic nucleus (AGN) can strongly affect its host. Due to the copious radiative power of the nucleus, the effects of radiative feedback can be detected over the entire host galaxy and sometimes well into the intergalactic space. In this paper we model the observed size-luminosity relationship of the narrow-line regions (NLRs) of AGN. We model the NLR as a collection of clouds in pressure equilibrium with the ionizing radiation, with each cloud producing line emission calculated by Cloudy. The sizes of the NLRs of powerful quasars are reproduced without any free parameters, as long as they contain massive (\SIrange{e5}{e7}{M_\sun}) ionization-bounded clouds. At lower AGN luminosities the observed sizes are larger than the model sizes, likely due to additional unmodeled sources of ionization (e.g., star formation). We find that the observed saturation of sizes at $\sim \SI{10}{kpc}$ which is observed at high AGN luminosities ($L_\text{ion}\simeq \SI{e46}{erg/s}$) is naturally explained by optically thick clouds absorbing the ionizing radiation and preventing illumination beyond a critical distance. Using our models in combination with observations of the \oiii/IR ratio and the \oiii\ size -- IR luminosity relationship, we calculate the covering factor of the obscuring torus (and therefore the type 2 fraction within the quasar population) to be $f=0.5$, though this is likely an upper bound. Finally, because the gas behind the ionization front is invisible in ionized gas transitions, emission-based NLR mass calculations underestimate the mass of the NLR and therefore of the energetics of ionized-gas winds.
\end{abstract}




\section{Introduction} \label{sec:intro}

Feedback from active galactic nuclei (AGN) is now thought to act as a key regulatory mechanism in galaxy formation. An AGN can influence its host through both mechanical and radiative feedback. Mechanical feedback in the form of quasar-driven outflows is of primary importance in regulation of galaxy formation \citep{hopkins2006}. The accretion energy of a supermassive black hole is more than sufficient to expel gas from the galaxy potential, thereby making AGN feedback a natural agent in cutting off star formation in massive galaxies \citep{silk1998,thoul1995,croton2006}. The observed relationship between black hole mass and stellar velocity dispersions is another possible indicator that the formation of black holes and of their host galaxies may be linked \citep{gebhardt2000,ferrarese2000}. Although many aspects of mechanical feedback are still poorly understood, in the last decade extensive observations across the electromagnetic spectrum have provided convincing evidence for this process in action \citep{greene2012,liu2013b,harrison2015,crichton2016,fiore2017}. 

Furthermore, powerful AGN are capable of ionizing the gas throughout the galaxy and well into the intergalactic space, providing radiative feedback onto the interstellar and intergalactic medium. Quasars, typically defined as AGN with bolometric luminosities $L_{\rm bol}>\SI{e45}{erg/s}$, have been associated with especially large-scale ionized gas structures. Recent deep searches have uncovered giant $\text{Ly}\alpha$ nebulae around quasars with sizes upwards of $\SI{100}{kpc}$ \citep{hennawi2013,hennawi2015,borisova2016}. Illumination of the intergalactic medium can have lasting effects on the physical state of the gas well after the end of the AGN episode \citep{oppenheimer2017}. Even moderate-luminosity AGN can ionize nearby galaxies, provided they fall within the unobscured paths to the nucleus \citep{keel2017}, and quasars ionize companion galaxies out to $\ga 100$ kpc \citep{liu2009, vill10}. Finally, photo-ionization by the AGN makes it possible to detect mechanical feedback, as the illuminated gas produces strong optical emission lines, in particular \oiiifull, making it possible to map its kinematics \citep{nesvadba2006, nesvadba2008, fabi12,liu2013b,harrison2014,brusa2016,carniani2016}.

In this paper we re-examine radiative feedback of quasars by modeling the narrow-line regions (NLRs) of type 2 quasars drawn from the samples identified in the Sloan Digital Sky Survey \citep{zakamska2003,reyes2008,yuan2016}. These objects are luminous and near enough for spatially resolved observations of the NLRs, and the obscuration of the nucleus allows for study of the NLRs without the glare of the central source. The standard approach to studying the NLRs is to use line ratios of an individual object to determine the physical parameters of the gas \citep{netzer1993,groves2004}. Here we take a somewhat different approach of analyzing the ensemble of NLRs, to gain additional insight into the physics of the extended gas using population data. 

One fundamental measurement is the size of the narrow-line region, which can be obtained from spectroscopic or narrow-band photometric observations. In an early study \cite{bennert2002} found that the sizes of the extended emission line regions increase with AGN luminosity as $R\propto L_\text{\oiii}^{0.52\pm 0.06}$. Interestingly, this scaling is reminiscent of the relationship derived from the reverberation mapping for the broad-line region (BLR) of AGN, $R\propto L^{0.6\pm 0.1}$ \citep{peterson2001}. The slope of the BLR relationship and the similarity of broad line ratios among AGN -- and therefore of the ionization parameter $U=\Phi/(4\pi r^2 n_e c)$, where $\Phi$ is the ionizing photon production rate -- suggest that the density distribution within the BLR is nearly independent of distance. The same arguments would apply to the NLR given the scaling found by \cite{bennert2002}, resulting in a puzzling requirement of a constant density of narrow-line clouds, in contrast with the well-known decline of density seen in starburst galaxies \citep{heck90}. Subsequent observations of type 2 Seyfert galaxies indicated a flatter relationship, $R\propto L_\text{\oiii}^{0.33\pm 0.04}$, which cannot be readily explained in the same way \citep{schmitt2003}. Complicating matters is the difference in the relationship for type 1 and type 2 AGNs, with type 1 AGNs exhibiting a steeper size-luminosity relationship \citep{bennert2006_1,bennert2006_2}. 

Comparisons among these results were impeded by the lack of a standard definition of a size of the emission-line regions. Recent sensitive integral field unit observations allowed for the NLR size to be defined in a sensitivity- and distance-independent way, as the isophotal radius $R_{15}$ corresponding to an \oiii\ surface brightness of $10^{-15}/(1+z)^4$ erg/s/cm$^2$/arcsec$^2$ \citep{liu2013}. Even more recent studies combine effects of mechanical and radiative feedback into the same definition \citep{sun2017}. Using these measurements, the relation between the NLR size and the luminosity in different bands has been studied more extensively. In particular, \citet{liu2013} find that despite the presumably wide range of gas masses and cloud properties within the population, AGN follow a rather flat ($R\propto L_{\rm [OIII]}^{0.25}$), relatively tight relationship between various luminosity measures and sizes over four decades in luminosity, with only 0.3 dex root-mean-square spread around the best-fit relationship \citep{sun2018}. The relationship is steeper when mapped as a function of infrared luminosity, but at high infrared luminosities ($L\approx \SI{5e44}{erg/s}$) sizes stop growing and saturate at $\sim 10$ kpc \citep{hainline2014}.

Here we build on calculations of \cite{stern2014,stern2016} to construct a theoretical model of the NLRs. Our goals are to model the observed size/luminosity relationships of NLRs and their multi-wavelength emission \citep{zakamska2005,greene2011,liu2013,greene2014,hainline2014,obied2016,sun2017}. We set up our model in Section \ref{sec:model}, present model results and comparison with observations in Section \ref{sec:results}, discuss the implications of our results and possible enhancements of the model in Section \ref{sec:discussion} and conclude in Section \ref{sec:conclusions}. Following the long-standing convention, the wavelengths of emission lines are given in air. Whenever necessary, we use an $h=0.7$, $\Omega_{\rm m}=0.3$, $\Omega_{\Lambda}=0.7$ cosmology. Lower-case $r$ typically indicates radius or distance in spherical coordinate systems, whereas upper case $R$ typically indicates distances projected onto the plane of the sky. 

\section{Model Setup}\label{sec:model}

\begin{figure}
\centering
\includegraphics[width=\linewidth]{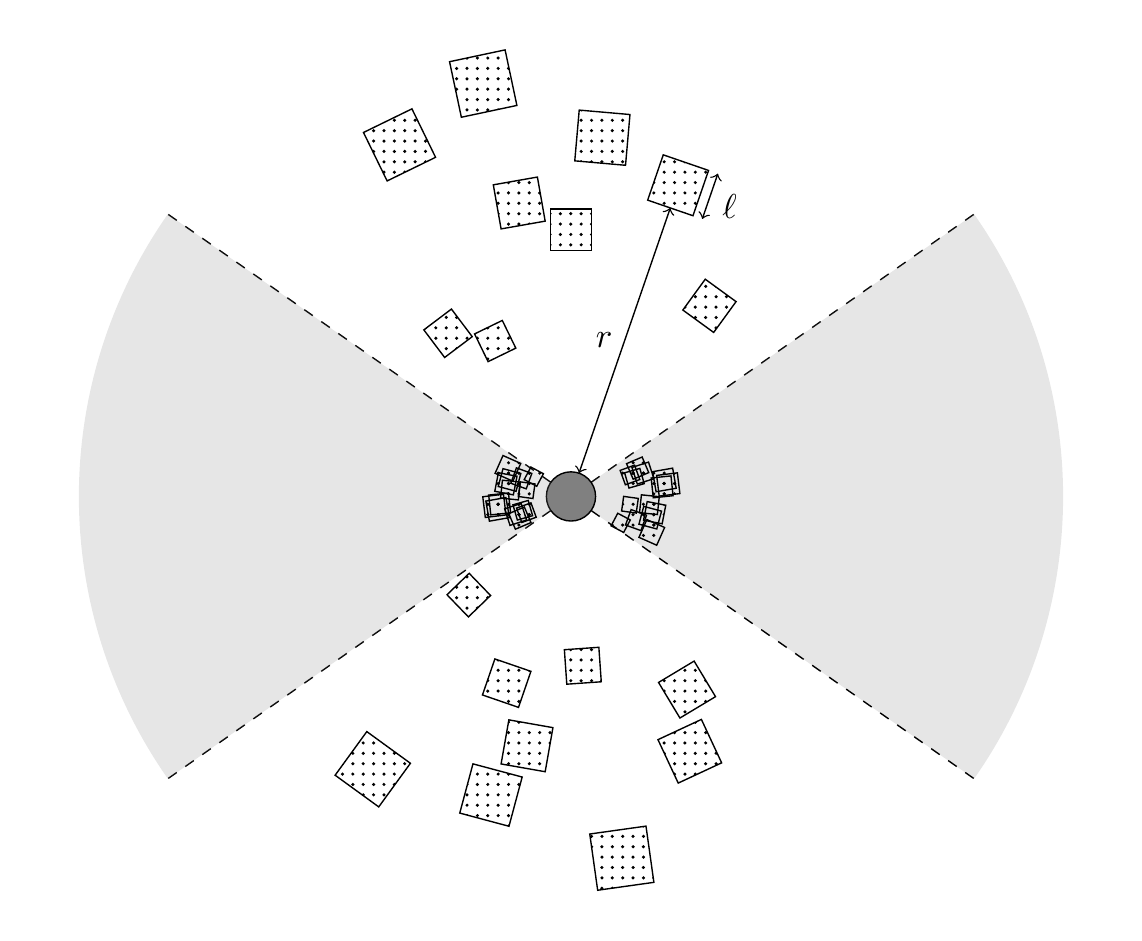}
\caption{We model the AGN as a central source surrounded by equatorial clouds which cover a solid angle $4\pi f$. The remaining solid angle is occupied by a bi-cone of volume-filling gas, embedded with dense clouds of optically thick photo-ionized gas. Shown here is a slice through the axis of symmetry in the AGN unification model.}
\label{fig:geometry_sketch}
\end{figure}

Our geometric model for an AGN is depicted in Figure \ref{fig:geometry_sketch}. We assume that the AGN produces ionizing photons isotropically. The surrounding material is a mix of low-density optically thin gas and higher density clouds. We then assume that density of clouds increases toward an equatorial plane to the point when they provide near-complete coverage of the central source, resulting in the same effect as an obscuring torus in the classical geometric unification model \citep{antonucci1993}. These equatorial clouds block all ionizing photons beyond $\sim \SI{1}{pc}$-scale distance, set by dust sublimation \citep{barv87}. They re-radiate thermally at near- and mid-infrared wavelengths, but do not contribute to the forbidden-line emission. 

We denote with $f$ the fraction of the quasar sky obscured by the equatorial clouds. The ionizing radiation from the AGN can escape along the directions within the remaining solid angle, illuminating sparser clouds that are further away from the central source (considered in Section \ref{sec:clouds}). These clouds have densities below the critical density for the forbidden-line emission, and the resulting NLR fills a bicone with its vertex at the central radiation source. We model photo-ionized \oiii\ emission in the illuminated region in Section \ref{sec:ionization}, and emission in the UV and X-ray bands in Section \ref{sec:multi_wavelength}. The torus itself is important in infrared emission, which we model in Section \ref{sec:ir_model}. 

\subsection{Cloud Structure}\label{sec:clouds}

Our model is based on an inhomogeneous gas distribution with significant clumping such that
the gas is organized into clouds. This is supported by many observations which indicate
the presence of a high-density component in the NLR. Emission line ratios such as 
[\ion{S}{2}]$\lambda$6716/$\lambda$6731\AA\ or [\ion{O}{2}]$\lambda$3729/$\lambda$3726\AA\
of the gas producing narrow-line emission show densities in excess of $\SI{100}{cm^{-3}}$
\citep{nesvadba2008}. Additionally, studies of starburst galaxies indicate the formation
of a two-phase interstellar medium in the presence of an outflow \citep{heckman2001}.

We assume the \oiii\ emitting gas cloud temperature is set by equilibrium with the ionizing
radiation of the AGN to be $\sim\SI{e4}{K}$. This is observationally supported by measurements of the [\ion{O}{3}]($\lambda$4959+$\lambda$5007)/$\lambda$4363\AA\ ratio \citep{bradley2004}. Since the temperature of the photo-ionized gas is nearly constant, the pressure in this phase is proportional to density \citep{heck90}.

To model the pressure of the cloud gas, we assume the ionization front of a cloud is supported
by radiation pressure from the central source \citep{dopita2002,stern2014,stern2016,davies2016}. Pressure 
equilibrium then implies that the pressure of a cloud at a distance $r$ from the source is given by
\begin{equation}
    P(r) = \frac{L}{4\pi c r^2},
\end{equation}
where $L$ is the ionizing luminosity of the quasar. Then, since the ionization front is isothermal at temperature $T$, the number density of particles within a cloud is given by
\begin{equation}\label{eq:gas_density}
    n_p(r) = \frac{P(r)}{kT} = \frac{L}{4\pi c k T r^2}.
\end{equation}
The major advantage of the radiation-pressure confinement model is that it naturally produces an ionization parameter which is independent of distance, explaining the similarity of emission line ratios among AGN \citep{dopita2002}. 

To model the distribution of the clouds themselves, we need two additional parameters, the
cloud mass $m_c$ and the covering factor $\Omega$ at a fiducial radius $r_s\equiv \SI{1}{kpc}$.
We start with a NLR consisting of clouds with a single mass $m_c$; in Section \ref{sec:o3_results} we generalize this to a distribution $d\Omega/dm$. Using a particle mass $m_p \equiv \SI{1.67e-24}{g}$, the
number of particles in a cloud is $m_c/m_p$, so the volume is
\begin{equation}
    V(r) = \frac{m_c/m_p}{n_p(r)} = \frac{4\pi c k T m_c r^2}{L m_p}.
\end{equation}
We treat the clouds as cubes with scale length $\ell(r)$, as is necessary for the Cloudy \citep{ferl98} calculations. The depth of a cloud is therefore
\begin{equation}\label{eq:cloud_size}
    \ell(r) = \left(V(r)\right)^{1/3} = \left(\frac{4\pi c k T m_c}{L m_p}\right)^{1/3} r^{2/3},
\end{equation}
and the cross-sectional area is
\begin{equation}
    A(r) = \ell(r)^2.
\end{equation}
If we define the scale area $A_s \equiv A(r_s)$, then $A(r) = A_s(r/r_s)^{4/3}$. In order to have
a covering factor $\Omega$ at $r_s$, the number of clouds at $r_s$ must be
\begin{equation}
    N_c(r_s) = \frac{4\pi r_s^2\Omega}{A_s},
\end{equation}
so the number density of clouds at $r_s$ is
\begin{equation}\label{eq:cloud_number}
    n_c(r_s) = \frac{N_c(r_s)}{4\pi r_s^2 \ell(r_s)} = \frac{\Omega}{A_s \ell(r_s)} = \frac{\Omega}{V(r_s)}.
\end{equation}

To completely determine the rest of our model for calculating the emission of the NLR, we now must set the spatial distribution of clouds. One physically motivated distribution is $n_c \propto r^{-2}$: such a distribution would arise if the radiative feedback on the clouds were accompanied by mechanical feedback in the form of an outflowing wind. As discussed in the introduction, radiative and mechanical feedback are distinct processes and radiative feedback is not necessarily accompanied by mechanical feedback. We start with this fiducial density profile and expand the model to include other profiles in Section \ref{sec:simplifications}. In the steady-state, constant velocity wind, equal mass of gas should be passing through any spherical surface centered on the AGN, which corresponds to an inverse-square density. For this cloud density profile, the area covered
by clouds at a distance $r$ is then proportional to $n_c(r)\cdot (r^2 \ell(r)) \cdot A(r) \propto r^2$, so the covering factor $\Omega$ is independent of distance.

\subsection{\oiii\ Ionization Profile}\label{sec:ionization}

\oiii\ emission from photo-ionized gas takes place between two distances from the nucleus. The minimum distance is regulated by collisional de-excitation of forbidden line emission above a certain critical density. We are focused here on \oiiifull\ emission, which has $n_\text{crit} = \SI{7e5}{cm^{-3}}$. The minimum distance from the nucleus at which \oiii\ can be produced is thus
\begin{equation}
    r_\text{min} = \sqrt{\frac{L}{4\pi c k T n_\text{crit}}}.
\end{equation}

The maximum distance is set by the transition of the clouds to the matter-bounded regime in which the entire cloud becomes ionized to higher ionization states, so that \oiii\ emission is very weak \citep{binette1996,liu2013}. We found in the previous section that $n_p(r)\propto r^{-2}$ and $\ell(r)\propto r^{2/3}$, so the column density scales as $N(r)\propto r^{-4/3}$. Since the column density decreases with distance from the nucleus, for every cloud mass there is some distance at which the optical depth of clouds is too low to support \oiii\ emission \citep{stern2014}.

The exact distance at which this occurs is set by a critical column density for the matter-bounded transition. The transition occurs when dust opacity becomes dominant over gas absorption, which occurs at an ionization parameter $U=10^{-2.2}$ \citep{netzer1993}. The corresponding matter-bounded column density is then
\begin{equation}
    N_\text{mb} = \frac{Uc}{R_H} = \SI{1.4e21}{cm^{-2}},
\end{equation}
where $R_H = \SI{1.4e-13}{cm^3/s}$ is the Case B recombination coefficient \citep{osterbrock2006}. In terms of this value, the distance beyond which the cloud is matter-bounded and the \oiii\ emission is strongly suppressed is
\begin{equation}\label{eq:matter_bounded}
    r_\text{break} = r_s\left(\frac{m_p A_s N_\text{mb}}{m_c}\right)^{-3/4}.
\end{equation}

For a first approximation we assume the gas clouds are sparse enough to not overlap, so all clouds lying between these distances are photo-ionized by the quasar to some depth. We explore corrections for high covering factors in Section \ref{sec:high_omega}. 

Photo-ionized clouds have a characteristic layer structure \citep{dopita2002}, in which higher ionization potential ions are produced closer to the ionizing source, whereas the back of the cloud can remain neutral (if the cloud is optically thick to the ionizing radiation and is therefore in the ionization-bounded regime). In the radiation-pressure dominated models \citep{groves2004}, the detailed structure of the photo-ionized cloud is calculated by \cite{stern2014} using Cloudy \citep{ferl98}. These authors tabulate \oiii\ line luminosities $L_{\text\oiii,0}(r)$ for an ionizing luminosity of $L=\SI{e45}{erg/s}$ and a covering factor $\Omega = 0.3$, so we have from this a luminosity density
\begin{equation}\label{eq:density}
    j_\text{\oiii}(r) = \frac{L_{\text\oiii,0}(r/\sqrt{L/\SI{e45}{erg/s}})}{r^2 \ell(r)}\frac{L}{\SI{e45}{erg/s}}\frac{\Omega}{0.3}.
\end{equation}
The \oiii\ flux is then given by the truncated Abel transform
\begin{align}\label{eq:flux}
\begin{split}
    F_\text{\oiii}(R) &= \int_{-\sqrt{r_\text{break}^2-R^2}}^{\sqrt{r_\text{break}^2-R^2}} j_\text{\oiii}(\sqrt{R^2+z^2})\,dz \\
    &= 2\int_{R}^{r_\text{break}} \frac{j_\text{\oiii}(r) r\,dr}{\sqrt{r^2-R^2}}.
\end{split}
\end{align}
The total \oiii\ luminosity can be computed using $L_\text{\oiii} = \int_{r_\text{min}}^{r_\text{break}} 2\pi R F_\text{\oiii}(R)\,dR$. In the conical model, in which the torus obscures a fraction $f$ of the luminosity from the central source, these quantities are additionally proportional to $(1-f)$.

The observed surface brightness in \si{erg/s/cm^2/arcsec^2} corresponding to this flux can be computed for redshift $z\ll 1$ using
\begin{equation}
    I(R) = \frac{F(R)}{4\pi(206265)^2}.
\end{equation}
It is appropriate to use the same formalism in observations for objects at any redshift, as long as the apparent nebular sizes are corrected for the cosmological surface brightness dimming. This is the method adopted by \cite{liu2013} and \cite{sun2017}. The surface brightness allows us to define the size of the nebula, $R_{15}$, such that $I(R_{15}) = \SI{e-15}{erg/s/cm^2/arcsec^2}$. Since the total mass is linearly divergent in our fiducial model, we define the nebula mass to be the mass of clouds within a sphere of radius $R_{15}$:
\begin{equation}
    M_{15} = \int_0^{R_{15}} 4\pi r^2 n_c(r) m_c \,dr.
\end{equation}

\subsection{Multi-wavelength Emission in NLR}\label{sec:multi_wavelength}

In addition to the \oiii\ line emission, we are interested in modeling other spectral features of the NLR. Here we give equations for the scattered UV luminosity and the X-ray luminosity. Both of these predictions are compared with observational data in Section \ref{sec:results}. 

The type 2 quasars we model have their intrinsic UV luminosity absorbed by the dusty torus. However, UV emission does reach the photo-ionized NLR, where it can be scattered by the dusty clouds toward the observer. UV light from the source is thus observable in scattered light even in type 2 objects \citep{antonucci1993}. We predict the total UV flux by assuming a Milky Way-like dust composition for the NLR clouds, and using a single-scattering approximation in which photons scatter off clouds and into the line of sight.

We start by computing the monochromatic UV luminosity at wavelength $\lambda$. Let $\sigma_\text{ext}$ denote the extinction cross section per H nucleon at this wavelength. Then for clouds at radius $r$, UV radiation penetrates to a depth
\begin{equation}
    d(r) = \frac{1}{\sigma_\text{ext} n_H(r)}.
\end{equation}
We compute UV scattering in spherically symmetric NLR (i.e., with a torus covering factor $f=0$), so anisotropy in individual scattering events are averaged out after integrating over the scattering clouds. We can therefore use the total scattering cross section $\sigma_\text{scat}$ to describe scattering by a single cloud. The probability of an incident photon being scattered by a cloud at radius $r$ is
\begin{equation}
    p(r) = \int_0^{d(r)} \left(1-\frac{x}{d(r)}\right)\sigma_\text{scat} n_H(r)\,dx.
\end{equation}
Since $d(r)\ll \ell(r)$ we can approximate $r$ as constant in the integral. The fraction $\xi$ of quasar light scattered by a cloud at radius $r$ is then
\begin{equation}
    \xi(r) = \frac{\ell(r)^2}{4\pi r^2} p(r) = \frac{\ell(r)^2}{8\pi r^2} \frac{\sigma_\text{scat}(\lambda)}{\sigma_\text{ext}(\lambda)}.
\end{equation}
For the cross sections $\sigma_\text{scat}(\lambda)$ and $\sigma_\text{ext}(\lambda)$, we use a carbonaceous-silicate dust grain model with a Milky Way size distribution and $R_V = 3.1$ \citep{weingartner2001,draine2003}.

To determine the total scattered UV luminosity, we integrate over the NLR and multiply by the monochromatic luminosity of the quasar at $\lambda$. This gives
\begin{equation}\label{eq:uv}
    \lambda L_{\lambda,\text{scat}} = \lambda L_\lambda \int_{r_\text{min}}^{r_\text{break}}\xi(r)n_c(r)\cdot 4\pi r^2\,dr.
\end{equation} 

In addition to the scattered UV light, the NLR also produces X-rays. The strong spatial correlation between the \oiii\ emission and the soft X-ray emission \citep{bianchi2006} suggests that both \oiii\ and X-rays originate in the same clouds. This correspondence is a key piece of the observational evidence for a single photo-ionized cloud structure in the NLR \citep{stern2014}. Working within this assumption, we can express the predicted X-ray luminosity from the highly-ionized surfaces of the NLR clouds in much the same way as we express the predicted \oiii\ luminosity from the lower ionization inner layer.

\cite{stern2014} calculated the extended X-ray luminosity values $L_{X,0}(r)$ in the band \SIrange{0.5}{2}{keV}, again for an ionizing luminosity of $L=\SI{e45}{erg/s}$ and a covering factor $\Omega = 0.3$. In exact analogy with equation (\ref{eq:density}), the extended X-ray luminosity density is
\begin{equation}\label{eq:xray_density}
    j_{X,\text{ext}}(r) = \frac{L_{X,0}(r/\sqrt{L/\SI{e45}{erg/s}})}{r^2 \ell(r)}\frac{L}{\SI{e45}{erg/s}}\frac{\Omega}{0.3}.
\end{equation}
The flux $F_{X,\text{ext}}(R)$ can be computed using equation (\ref{eq:flux}).

\subsection{Infrared Emission in Conical Model}\label{sec:ir_model}

Type 2 quasars are strong emitters at mid-infrared wavelengths, but this emission is unlikely to originate in the NLR. The temperature of dust particles subjected to direct radiation from the nucleus decreases as roughly $T\propto r^{-1/2}$. The densely packed clouds in the equatorial region which constitute the obscuring torus are as close as $\sim \SI{1}{pc}$ from the AGN, are heated to temperatures of hundreds of K, and radiate at 2$-$10\micron, making AGN quite luminous at these wavelengths \citep{richards2006}. Therefore, in the classical unification model the mid-infrared emission originates from the obscuring torus, and the gas in the shadow of the torus does not produce \oiii\ emission. In turn, in the unshadowed region the sparsely distributed NLR clouds are too far from the quasar to have appreciable mid-infrared emission. 

\cite{hainline2014} and \cite{sun2017} present the relationship between \oiii\ sizes and infrared luminosities of type 2 quasars, which we seek to model here. Radiative transfer through the torus is a complex problem \citep{pier1993,nenkova2002,honi06}, where theoretical models are not yet always in quantitative agreement with observations. Instead of using such models we adopt a phenomenological approach and introduce a parameter $\varepsilon$ indicating the fraction of intercepted luminosity which is eventually emitted in the IR. In principle, $\varepsilon$ is wavelength-dependent, but for consistency with \cite{hainline2014} we are specifically interested in IR luminosities $\nu L_\nu$ at rest-frame $8\micron$. Therefore, in this paper $\varepsilon$ encompasses all the complexities of thermal emission efficiency and radiative transport at this particular wavelength of interest. The radiation emitted from the central source is essentially all at ionizing energies, so we can use the ionizing luminosity $L$ as a proxy for the bolometric luminosity entering the torus. The IR luminosity $\nu L_{\nu}$[8\micron] is then
\begin{equation}\label{eq:ir}
    L_{\text{IR}} = Lf\varepsilon.
\end{equation}
AGN tori are likely optically thick at mid-infrared wavelengths, and therefore even at these wavelengths the AGN emission cannot be considered isotropic. This is evidenced by obscured AGN having redder mid-infrared colors than the unobscured ones \citep{hickox2017}. Our parametrization in equation (\ref{eq:ir}) does not take into account this complexity. 

\section{Model Results}\label{sec:results}

The simplest case for our model is in the absence of a dusty torus, such that all emission comes from
NLR clouds. Then $f = 0$ and $\varepsilon$ is irrelevant, so there are only three free parameters: 
the ionizing quasar luminosity $L$, the mass of each cloud $m_c$, and the covering factor at a fiducial distance of $\SI{1}{kpc}$, $\Omega$ (which in the case of a $r^{-2}$ cloud density profile is independent of distance).

\subsection{\oiii\ emission in NLR}\label{sec:o3_results}

The gas clouds in the NLR are photo-ionized and produce the \oiii\ emission lines which we observe in extended emission-line region studies. With the model outlined in Section \ref{sec:model}, we can now predict several different observables associated with the \oiii\ measurements. In the following plots we use the parameter ranges $\SI{e5}{M_\sun}<m_c<\SI{e7}{M_\sun}$ and $\num{3e-4}<\Omega<\num{3e-2}$. These are chosen to provide reasonable integrated constraints, namely the total \oiii\ luminosity and the mass $M_{15}$, with observations. Details of the comparison of $M_{15}$ with other mass estimates are given in Section \ref{sec:mass_discussion}.

\begin{figure}
\includegraphics[width=\linewidth]{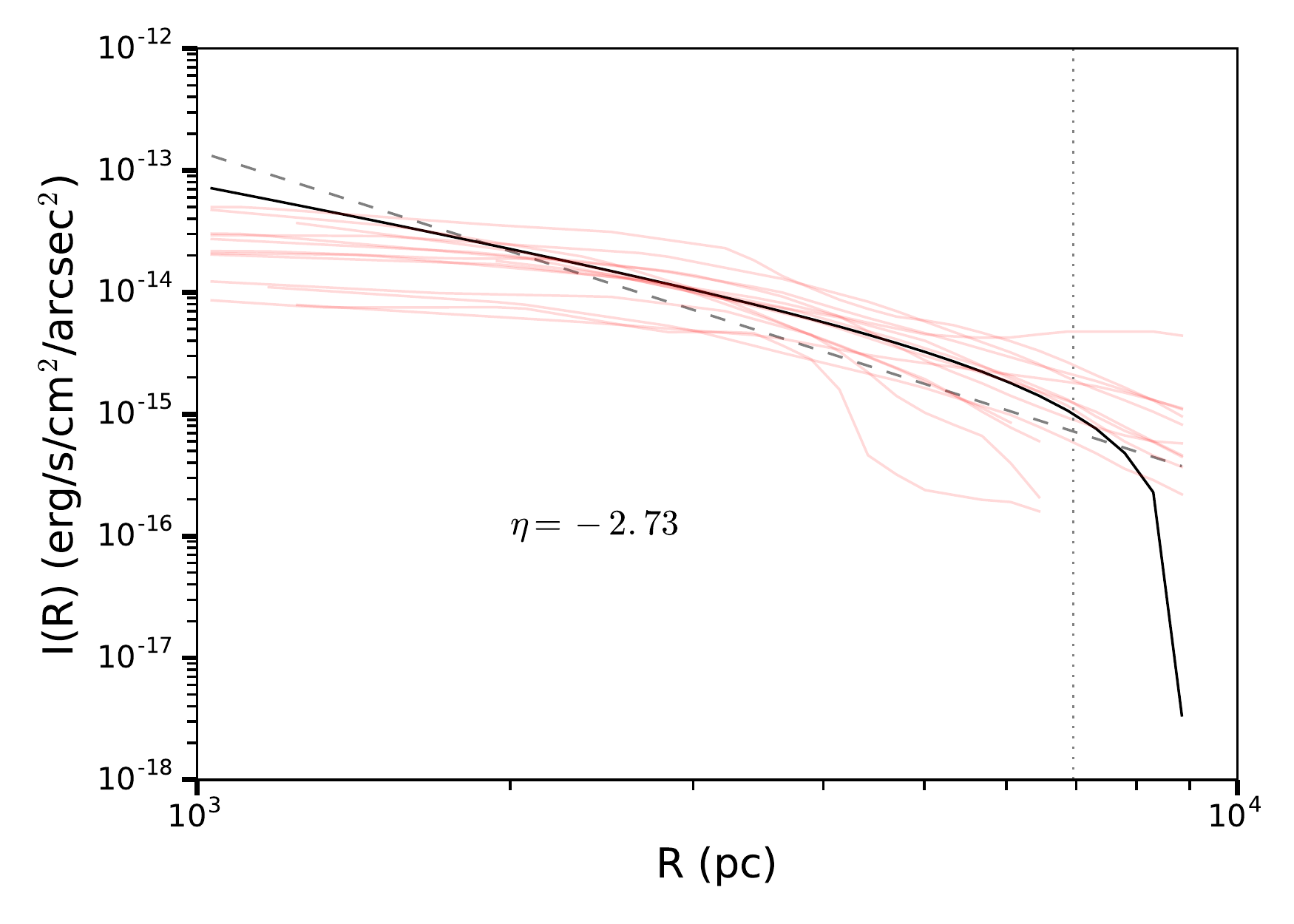}
\caption{The predicted observable surface brightness for $L=\SI{e45}{erg/s}$, $m_c = \SI{e6}{M_\sun}$, and $\Omega = \num{3e-3}$. The dashed line is the linear best fit with a slope $\eta = -2.73$, and the dotted axis is at $R=R_{15}$. The red curves are surface brightness profiles of luminous quasars at $z\sim 0.5$ \cite{liu2013}.\label{fig:surface_brightness}}
\end{figure}

One of the observables that recently became available with spatially resolved long-slit or integral-field observations is the surface brightness of the \oiii\ emission as a function of distance from the quasar. In Figure \ref{fig:surface_brightness} we show the surface brightness prediction of our model for the case of $m_c=\SI{e6}{M_\sun}$ and $\Omega=\num{3e-3}$. In this case the surface brightness reaches $\SI{e-15}{erg/s/cm^2/arcsec^2}$ at $R_{15} = \SI{5.5}{kpc}$. The profile is approximately a power law $\propto R^{-2.5}$ with a sharp cutoff at $r_\text{break} = \SI{8.9}{kpc}$, where the column density in the clouds drops below $N_\text{mb}$, the clouds become matter-bounded and produce little \oiii\ emission. The presence of matter-bounded clouds in the NLR was suggested by \cite{binette1996} and \cite{liu2013b}, and we explore the implications of the presence of matter-bounded clouds here. From equations (\ref{eq:cloud_size}) and (\ref{eq:matter_bounded}), we have $r_\text{break}\propto m_c^{1/4}$, so detection of the matter-bounded transition in observations provides a measurement of the maximal cloud mass.

The observations suggest a matter-bounded transition at $7.0\pm\SI{1.8}{kpc}$ \citep{liu2013}. This is in rough agreement with our prediction of $r_\text{break} = \SI{8.9}{kpc}$ obtained for fiducial $10^6 M_{\odot}$ clouds; less massive clouds undergo this transition at smaller distances from the AGN in agreement with equation (\ref{eq:matter_bounded}). However, the predicted slope of the surface brightness profile does not agree well with observations. We measure the predicted slope using data upwards of $\SI{1}{kpc}$, to account for the point spread function of the observations which induces approximately this resolution. Our model gives a slope of $\eta = -2.73$, which is shallower than the observed values of $3<\eta<6$ \citep{liu2013}.

\begin{figure}
\includegraphics[width=\linewidth]{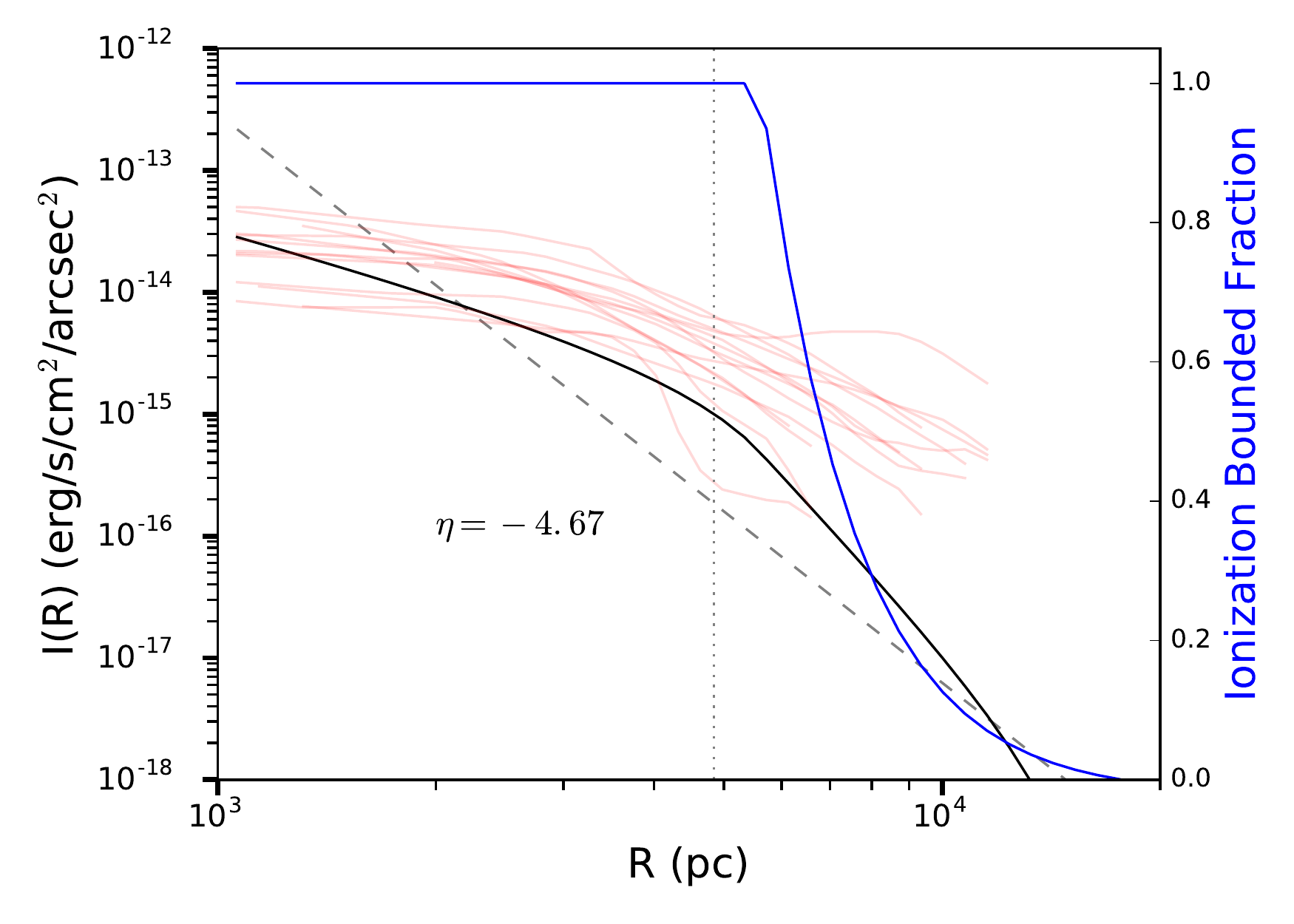}
\caption{The predicted observable surface brightness for $L=\SI{e45}{erg/s}$ with a total covering factor $\Omega = \num{3e-3}$ and a power law mass distribution set by observations of Milky Way molecular clouds. The dashed line is the linear best fit with a slope $\eta = -4.67$. The fraction of ionization-bounded clouds is shown in blue. The red curves are surface brightness profiles of luminous quasars at $z\sim 0.5$ \citep{liu2013}.\label{fig:mass_distribution}}
\end{figure}

The difference in slopes may be an artifact of our assumption in Section \ref{sec:clouds} of $\d{\Omega}{m} = \Omega\delta(m-m_c)$. We can generalize the results of Section \ref{sec:clouds} to allow for a different distribution of cloud masses. In this case, clouds of different masses become matter-bounded at different radii (eq. \ref{eq:matter_bounded}), which would steepen the surface brightness profile. Figure \ref{fig:mass_distribution} shows an \oiii\ surface brightness profile for a covering factor $\Omega = \num{3e-3}$ and a power-law mass distribution
\begin{equation}
    \d{N}{\log m_c} = N_\text{cu}\left(\frac{m_u}{m_c}\right)^{-(\alpha_M-1)}\quad\text{for}\quad m_c\le m_u,
\end{equation}
where $N_\text{cu}$ is set by $\Omega$ in our model, $m_u = \SI{6e6}{M_\sun}$, and $\alpha_M = 1.7$. This mass distribution has been observed for giant molecular clouds in the Milky Way \citep{williams1997, nakanishi2016, miville2017}.

Figure \ref{fig:mass_distribution} shows that the matter-bounded transition takes place over a range of several kiloparsecs as the ionization-bounded fraction decreases to zero. Moreover, this transition is responsible for increasing the slope to $\eta = -4.67$, which is a typical value in comparison with observed profiles \citep{liu2013}. However, alterations in the mass distribution $\d{\Omega}{m}$ produce the same effects on surface brightness as alterations in the cloud density profile $n_c(r)$, since both parameters change the number of \oiii-emitting clouds at a given distance. Different cloud density profiles could produce this surface brightness slope, as well as other predictions which are explored further in Section \ref{sec:simplifications}.

In addition to raw observations of surface brightness profiles, spatially resolved integral-field observations have permitted the development of a size metric for the NLR which does not depend on redshift or depth of observations, the photometrically defined distance $R_{15}$. Measurements of this size in samples of AGN have been found to correlate strongly with the \oiii\ luminosity \citep{liu2013}. We can construct a prediction for the size-\oiii\ luminosity relationship using the surface brightness profiles predicted by our model for a range of ionizing luminosity values.

\begin{figure}
\includegraphics[width=\linewidth]{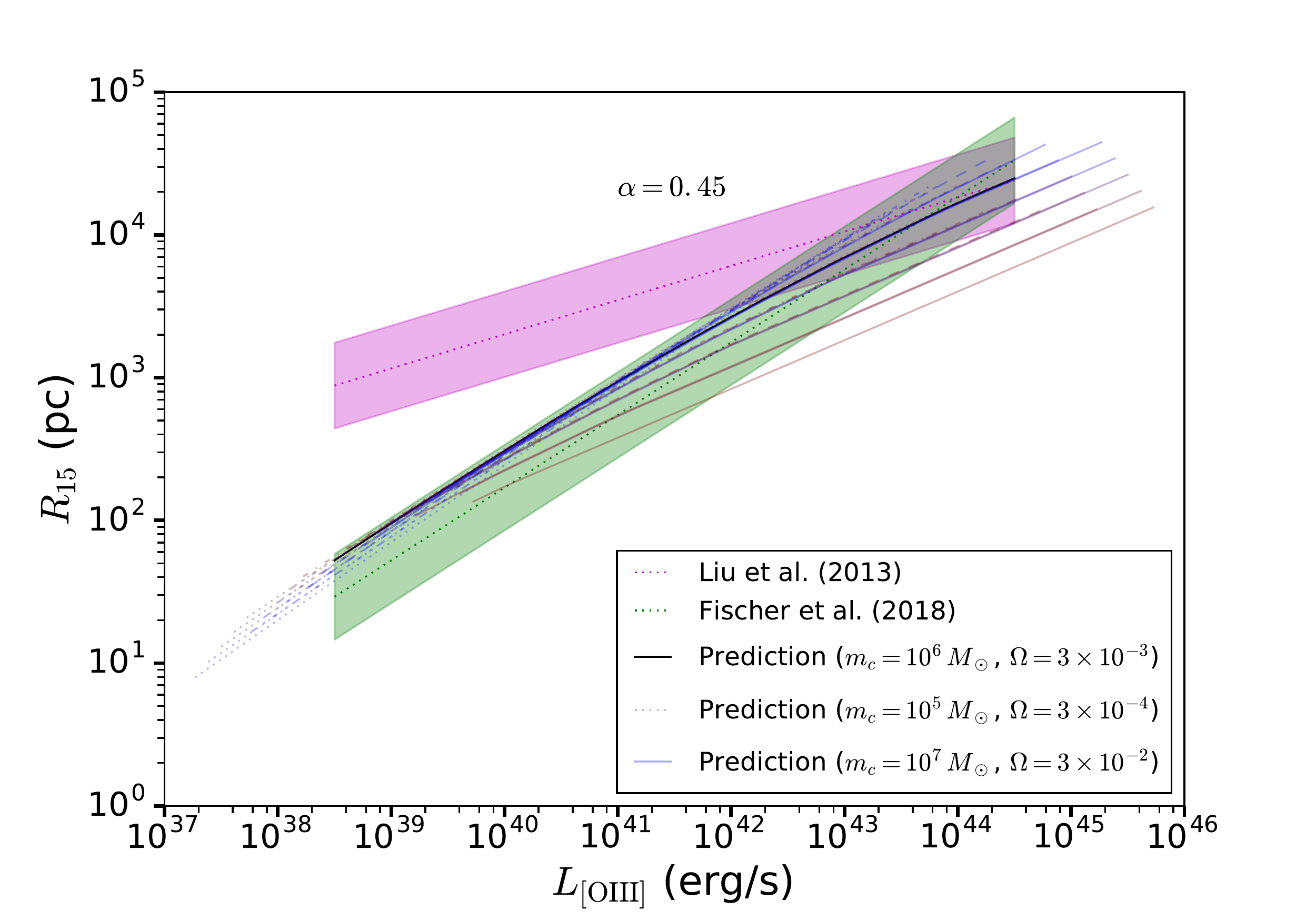}
\caption{The size-\oiii\ luminosity relationship for quasars with source luminosities in the range \SIrange{e42}{e46}{erg/s} using parameters $m_c = \SI{e6}{M_\sun}$ and $\Omega = \num{3e-3}$. The best fit line has a slope $\alpha = 0.48$. Also plotted in dotted lines is the same relationship for values sampled from $\SI{e5}{M_\sun} \le m_c \le \SI{e7}{M_\sun}$ and $\num{3e-4} \le \Omega \le \num{3e-2}$. Higher $m_c$ models are shown in blue, and higher $\Omega$ models are shown in thicker lines. The observed relationship from \cite{liu2013} is shown in the purple region, with dispersion 0.3 dex \citep{sun2018}. Another size-luminosity fit\replaced{, obtained by replacing points from \citet{fraquelli2003} with points from \citet{schmitt2003},}{ from \citet{fisc18}} is shown in the green region. \label{fig:size_luminosity}}
\end{figure}

Figure \ref{fig:size_luminosity} shows our prediction in comparison with the observations \replaced{\citep{liu2013}}{\citep{liu2013,fisc18}}, using a range of parameter values within \SI{1}{dex} of the fiducial values $m_c = \SI{e6}{M_\sun}$ and $\Omega = \num{3e-3}$. The most striking feature of Figure \ref{fig:size_luminosity} is that as long as the mass of the clouds is high enough, the models form a narrow locus in the size-luminosity space (superposed blue and grey curves). The position of objects along the locus is driven by $L$ and $\Omega$, which affect the resulting \oiii\ luminosity in an identical way (eq. \ref{eq:density}). The physical reason for this behaviour is that piling additional mass behind the ionization front in ionization-bounded clouds results in no changes to the NLR emission. (We have verified that models with a broad cloud mass distribution $\d{\Omega}{m_c}$ also follow the locus as long as the distribution extends to high enough masses.) Therefore, in the presence of ionization-bounded clouds in the NLR the size-luminosity relation is observable even if there is wide variation throughout the AGN population in the cloud mass or cloud covering factor. 

Another encouraging feature of Figure \ref{fig:size_luminosity} is that the model size-luminosity relations with values of $m_c\ga 10^6$ M$_{\odot}$ are in excellent quantitative agreement with those seen by \citet{liu2013} at high luminosities $L_{\rm [OIII]}=10^{44}$ erg s$^{-1}$. Aside from the requirement to have some high-mass ionization-bounded clouds and our fixed inverse-square $n_c(r)$ profile, the model for the size-luminosity relationship has no free parameters: the sizes of the NLR are completely constrained by the photo-ionization physics. Unlike the surface-brightness profiles, which vary strongly depending on the cloud mass distribution due to the presence of the matter-bounded clouds, the size-luminosity relationship is much less sensitive to $\d{\Omega}{m}$, as long as some ionization-bounded clouds are present. 

The most notable discrepancy between our models and the observations is that our predicted slope of $\alpha = 0.45$ is higher than the observed value of $0.25\pm 0.02$ \citep{liu2013}. Observations conducted by \citet{liu2013} directly constrain the size-luminosity relationship at high $L_{\rm [OIII]}$, where our models show quantitative agreement with the data. In contrast, at lower luminosities \citet{liu2013} took archival measurements \citep{fraquelli2003} of sizes and re-cast them in terms of $R_{15}$ with some assumptions, which could have introduced additional uncertainties in the faint-end sizes and in the resulting slopes. Recent measurements by \citet{sun2018} map the size-luminosity relationship directly over several decades of luminosity and find a slope of 0.3, slightly steeper than \citet{liu2013}, but shallower than those seen in our model. However, the \citet{sun2018} sample is selected based on [OIII] line luminosities, so the measured slope may be biased.

\deleted{Due to this uncertainty in the low-luminosity end of the observed size-luminosity relationship, we have also included a slope which uses archival measurements of \citet{schmitt2003} rather than \citet{fraquelli2003}. \citet{schmitt2003} measured surface brightness of \oiii\ down to a 3$\sigma$ cutoff of about \SI{3.7e-15}{erg/s/cm^2/arcsec^2}. We assume a surface brightness profile proportional to $R^{-3}$, and thus increase the reported sizes by a factor of $a^{1/3}$, where $a$ is the $3\sigma$ cutoff in units of \SI{e-15}{erg/s/cm^2/arcsec^2}, in order to estimate $R_{15}$. We additionally increase all reported sizes by 0.2 dex to account for seeing \citep{hainline2014}. By combining the adjusted Schmitt points with other data in the \citet{liu2013} sample, we obtain a power law fit with a slope of 0.43. This agrees much more closely with our model. However, since there is significant uncertainty associated with the method used to obtain $R_{15}$ values from the \citet{schmitt2003} measurements, it is not clear to us that this method of measuring the slope of the size-luminosity relationship is more reliable than the others reported in the literature. Rather, we consider the slope to be uncertain but likely in the range 0.25--0.45.}

\added{Due to this uncertainty in the low-luminosity end of the observed size-luminosity relationship, present here another size-luminosity relationship closely following \citet{fisc18}, who use archival measurements of \citet{schmitt2003} rather than \citet{fraquelli2003} on the low-luminosity end. \citet{fisc18} jointly analyze intermediate-luminosity type 2 objects, high-luminosity type 2 quasars presented by \citet{liu2013}, and type 1 and 2 Seyfert galaxies from \citet{schmitt2003}. With all of these objects they find a slope of 0.52; we remove the type 1 Seyfert galaxies since our model applies only to type 2 objects, leading to the final slope of 0.51. This slope agrees more closely with our model than that of the relationship in \citet{liu2013}. However, since there is intrinsic uncertainty in the low-luminosity measurements, it is not clear to us which value of the slope of the size-luminosity relationship is the most reliable. Rather, we consider the slope to be uncertain but likely in the range 0.25--0.5.}

The most likely explanation for the discrepancy between the model and in some of the low-luminosity observations is the presence of additional sources of ionization on the faint end of the size-luminosity relationship. For example, extended star formation in the host galaxy can result in relatively low surface-brightness emission of ionized gas over the entire galaxy which could be captured by the $R_{15}$ definition and artificially inflate the apparent NLR sizes. \citet{sun2018} find that a higher surface brightness limit may potentially be a better measure of NLR sizes. This effect is less important in powerful quasars which dominate photo-ionization, explaining why the model produces a good quantitative agreement with the data in that range. 

The self-similarity does break down at higher luminosities. This is due to the matter-bounded transition occurring at different distances for different cloud masses. Since lower cloud masses go through the matter-bounded transition at lower distances, they reach an upper size limit more quickly, beyond which the size-luminosity slope is lower due to the lack of \oiii\ emission at distances beyond this limit. If we incorporated a physically realistic mass distribution $\d{\Omega}{m}$, then the matter-bounded transition would be spread over a range of distances, as in Figure \ref{fig:mass_distribution}. The resulting size-luminosity curve would be in the region covered by the family of curves in Figure \ref{fig:size_luminosity} (assuming the cloud masses are in the range $\SI{e5}{M_\odot}<m_c<\SI{e7}{M_\odot}$). In order for the prediction to agree with the observations of \cite{liu2013}, the mass should be at least $\SI{e6}{M_\odot}$.

However, changing the density profile from $n_c \propto r^{-2}$ to some other dependence on $r$ can have a significant effect on the size-luminosity relationship. This is because $\Omega(r)$ ceases to be constant, which leads to a different dependence of $L_\text{\oiii}$ on $L$. At the same time, changing the density profile has roughly the same effect on the surface brightness as changing the mass distribution. Thus, $R_{15}$ is not substantially affected while $L_\text{\oiii}$ is, leading to a change in the size-luminosity slope. The potential of this effect to explain the observed slope in \cite{liu2013} is explored in Section \ref{sec:simplifications}.

\subsection{UV and X-ray emission from illuminated NLR}\label{sec:multi_wavelength_results}

Circumnuclear obscuration prevents all of the direct UV light from the AGN from reaching the observer in type 2 objects. Nonetheless it has long been known that type 2 Seyferts and quasars produce an appreciable amount of UV emission \citep{antonucci1993, zakamska2003}. While in some cases this emission is likely due to star formation in the host galaxy occurring contemporaneously with AGN activity \citep{heckman1997}, UV imaging and polarimetry clearly indicate that much of it arises in conically shaped regions, likely due to scattering of the quasar light off the material in the interstellar medium and into the observer's line of sight \citep{hines1999,zakamska2005,zakamska2006,obied2016,wylezalek2016}. It is not known which of the components of the illuminated interstellar medium dominate scattering, though initial estimates of scattering efficiency suggested that scattering is dominated by gas with densities that are much lower than those of the NLR \citep{greene2011, obied2016}. 

To determine the typical contribution of NLR scattering of the UV emission to the observed UV light, we first establish the typical observed amount of UV emission. To this end, we cross-match type 2 quasars from \citet{reyes2008} and \citet{yuan2016} to the Galaxy Evolution Explorer (GALEX) data archive \citep{morrissey2007,bianchi2014}.  We use a subsample from $0.1<z<0.2$ to minimize redshift dependence (including the redshift dependence of k-corrections) and to maximize the number of UV detections, given the relatively low sensitivity of GALEX observations. There are 238 such objects in \citet{reyes2008} and 46 in \citet{yuan2016}. Within this redshift range, 65\% of objects are detected in the far-UV band (1516\AA) and 81\% are detected in the near-UV band (2267\AA), and we use upper limits on fluxes for the rest. As our model is bench-marked to the total \oiii\ emission, and in the illumination model UV and \oiii\ luminosities are expected to scale the same way with the model parameters, the most useful measurement is that of the UV-to-\oiii\ ratio. We obtain the median observed ratios of
\begin{align*}
    \frac{\lambda L_{\lambda,\text{ext}} (1516\,\text\AA)}{L_{\text\oiii}} &= 15.2,\\
    \frac{\lambda L_{\lambda,\text{ext}} (2267\,\text\AA)}{L_{\text\oiii}} &= 13.6,
\end{align*}
and the sample standard deviation around these relationships is $\sim 0.4$ dex. 

To compare these values with our model predictions, in Section \ref{sec:multi_wavelength} we compute the monochromatic scattered UV luminosity at a wavelength $\lambda$. This value is proportional to the monochromatic luminosity of the quasar at $\lambda$. We use a composite quasar spectral energy distribution from \citet{richards2006} to calculate the ratio of $\lambda L_\lambda$ to the ionizing luminosity $L$ (obtained by counting emission at photon energies above $\SI{13.6}{eV}$). We can then use $L$ to compute the \oiii\ luminosity, and $\lambda L_\lambda$ to compute the scattered UV luminosity. The predicted ratios at fiducial wavelengths $1516\,\text\AA$ and $2267\,\text\AA$ are
\begin{align*}
    \frac{\lambda L_{\lambda,\text{scat}} (1516\,\text\AA)}{L_{\text\oiii}} &= 2.76,\\
    \frac{\lambda L_{\lambda,\text{scat}} (2267\,\text\AA)}{L_{\text\oiii}} &= 2.83.
\end{align*}
This suggests that 10-20\% of the observed extended UV flux originates from scattering in the NLR clouds, and thus is observable even with total obscuration of the central source. This confirms the previously held assumption that the majority of scattered UV light originates in a low-density volume filling gas \citep{zakamska2005,greene2011,obied2016}.


We can similarly compare our predicted X-ray luminosities with observations. An important mechanism accounting for soft X-rays from the NLR is photo-ionized line emission from regions of the NLR clouds at high ionization states \citep{greene2014}. Indeed, recombination lines observed in the spatially resolved spectra of NGC 4151 strongly suggest that a photo-ionized component of the NLR contributes significantly to the extended X-rays \citep{ogle2000}. We can thus use the predicted X-ray emission from the photo-ionized NLR clouds to estimate the extended X-ray luminosity.

The luminosity of X-rays in the \SIrange{0.5}{2}{keV} band due to photo-ionization is given in equation (\ref{eq:xray_density}), which is analogous to the \oiii\ luminosity given in equation (\ref{eq:density}). Therefore, our model predicts an X-ray/\oiii\ ratio which, like the UV/\oiii\ ratio, is independent of all the model parameters. Its value is $L_X/L_\text{\oiii} = 0.04$, which is set by the ratio of the \oiii\ and X-ray luminosities in photo-ionization models of NLR clouds \citep{stern2014}.

Spatially resolved X-ray studies of Type 2 AGN have generally suggested roughly constant X-ray/\oiii\ ratios in the range $0.1 < L_X/L_\text{\oiii} < 0.3$ \citep{wang2011,paggi2012,greene2012}. The discrepancy between our model value and these observations is likely due to additional sources of extended X-rays. This provides independent support to observations which have also indicated the presence of extended X-ray emission from sources other than photo-ionized gas \citep{bianchi2006}. The source may be soft X-ray emission from a hot gas in collisional equilibrium, although the astrophysical nature of this gas remains a matter of speculation \citep{bianchi2010}.


\subsection{Infrared emission from illuminated torus}\label{sec:ir_results}

\begin{figure*}[t]
\includegraphics[width=\linewidth]{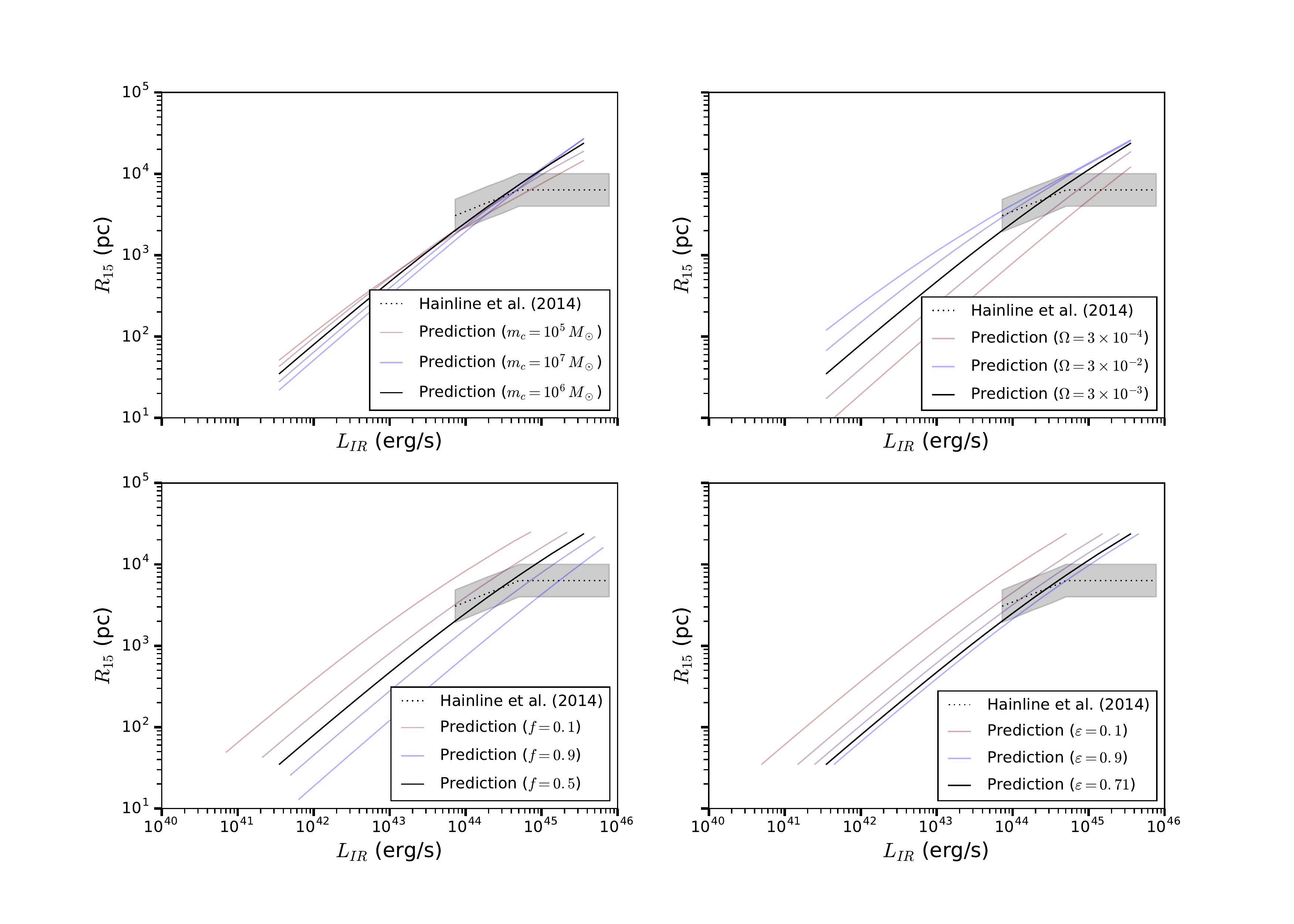}
\caption{The size-IR luminosity relationship for quasars with source luminosities in the range \SIrange{e42}{e46}{erg/s} using parameters $m_c = \SI{e6}{M_\sun}$, $\Omega = \num{3e-3}$, $f=0.5$, and $\varepsilon = 0.71$. The best fit line has a slope $\beta = 0.72$. Each panel shows the result of adjusting one of the parameters within the range given in the legend. Also plotted is the observed size-IR luminosity relationship \citep{hainline2014}. \label{fig:size_ir_luminosity}}
\end{figure*}

When we include the dusty torus, there are two additional model parameters: the covering factor of the torus $f$ and the proportion of light absorbed by the torus which is reradiated in the infrared, $\varepsilon$. The value of $f$ is tied to the relative frequency of Type 1 and Type 2 AGNs, whereas the value of $\varepsilon$ is set by the luminosity ratio $L_\text{torus}/L$. \citet{treister2008} derive the constraints on these parameters using infrared observations of type 1 AGN. In a similar spirit, in Section \ref{sec:ir_parameters} we use the infrared observations of type 2 AGN and the \oiii\ size -- IR luminosity relationship \citep{hainline2014} to set constraints on $f$ and $\varepsilon$. In the meanwhile, for our fiducial models of the size-luminosity relationship we use the values $f=0.5$ and $\varepsilon = 0.71$. 

With these parameters set, we can compute the IR luminosity and the size $R_{15}$ as a function of ionizing luminosity $L$. The relationship is plotted in Figure \ref{fig:size_ir_luminosity}, along with the dependence of the relationship on each of the model parameters. The observed relationship in Type 2 AGN is shown in each panel for comparison \citep{hainline2014}. At $L_{\rm IR}=10^{44.5}$ erg s$^{-1}$, our fiducial model quantitatively reproduces the observed sizes of the NLRs of AGN. At lower luminosities, the slope of the model relationship ($\sim 0.7$) is steeper than the slope of the observed relationship ($\sim 0.5$), as was also the case for the size-\oiii\ luminosity relationship in Section \ref{sec:o3_results}. As already discussed in that section, at lower luminosities additional sources of ionization may be increasing the apparent NLR sizes beyond what can be produced by the AGN. Although our models do not capture this regime quantitatively, we find that the models successfully reproduce the difference in slopes between the size-\oiii\ luminosity and size-IR luminosity relationship: the latter are steeper than the former, with power-law index difference of $0.2-0.3$ in both observations and in our models.

Although the models do show a very slight flattening at at higher luminosities due to the matter-bounded transition, none of the predicted curves exhibit the sharp transition to a flat relationship at a maximum size of $\la 10$ kpc \citep{hainline2014}. The weak dependence of the relationship on cloud mass, shown in the upper left panel of Figure \ref{fig:size_ir_luminosity}, indicates that the matter-bounded transition is likely insufficient to explain the transition in slope. Another possible explanation for the size cutoff is explored in Section \ref{sec:high_omega}.

\section{Discussion}\label{sec:discussion}

\subsection{Robustness of the Size-Luminosity Relationship}\label{sec:robustness}

One of the most striking predictions of this model is the stability of the size-\oiii\ luminosity relationship over a wide range
of the parameters $\Omega$ and $m_c$, at luminosities where the matter-bounded transition is not in effect. This has a simple physical interpretation. The \oiii\ luminosity density in Equation
\ref{eq:density} is proportional to the product $L\Omega$. Thus, the effects of varying $\Omega$ on $L_{\text{\oiii}}$ and 
$R_{15}$ are the same as the effects of varying the source luminosity $L$. Since the size-luminosity relation is computed 
by varying $L$, a change in $\Omega$ only shifts points along the same curve, without altering the curve itself.

The insensitivity to $m_c$ is a result of the NLR clouds being optically thick to ionizing radiation for $r<r_\text{break}$. The \oiiifull\ light is emitted at the ionization front of the clouds. Thus, increasing $m_c$ only changes the amount of gas behind the ionization front, which has no effect on the \oiii\ emission. This argument also shows why there is some residual dependence on $m_c$: since  $r_\text{break}\propto m_c^{3/4}$, an increase in $m_c$ results in more clouds becoming ionization-bounded and contributing to the \oiii\ luminosity.

\subsection{Flattening of the Size-Luminosity Relationship}\label{sec:high_omega}

A key property of the quasar size-luminosity relationship is its flattening at high luminosities \citep{hainline2014}. We 
propose two potential physical explanations for this phenomenon on the basis of our cloud model. 

One explanation is the dependence of the matter-bounded transition on cloud mass. Since $r_\text{break}\propto m_c^{3/4}$,
lower mass clouds become matter-bounded at lower radii, so the surface brightness becomes steeper at larger $R$. Thus,
at higher luminosities, further increases in $L_\text{\oiii}$ have less of an effect on $R_{15}$, leading to a flattening
of the size-luminosity relationship. This is most clearly observed when modeling a cloud population with a range of masses,
as in Figure \ref{fig:mass_distribution}.

Another phenomenon which may be related to the flattening is the saturation of the quasar sky with optically thick clouds. 
We have assumed in our calculations that clouds do not overlap, so every cloud with $r_\text{crit} < r < r_\text{break}$
is photo-ionized and contributes to \oiii\ emission. This assumption is only valid when the total covering factor,
\begin{equation}
    \Omega_\text{tot} = \int_{r_\text{crit}}^{r_\text{break}} \frac{\Omega(r)}{\ell(r)}\,dr,
\end{equation}
is much less than 1. Previous studies have reported $0.02\lesssim \Omega_\text{tot} \lesssim 0.2$, substantiating this assumption \citep{baskin2005}.  However, in our model with $m_c = \SI{e6}{M_\sun}$ and $\Omega = \num{3e-3}$, $\Omega_\text{tot}$ reaches a maximum of 0.88 for $L = \SI{e46}{erg/s}$, violating this condition. This means that at high luminosities, some of the available NLR clouds are shielded by optically thick clouds of lower radii, and thus do not participate in \oiii\ emission. The effect is qualitatively similar to the effect of the matter-bounded transition: at larger radii, fewer clouds are contributing  to \oiii\ emission, leading to a steepening of the surface brightness profile and a corresponding flattening of the size-luminosity relationship.

\begin{figure}
\includegraphics[width=\linewidth]{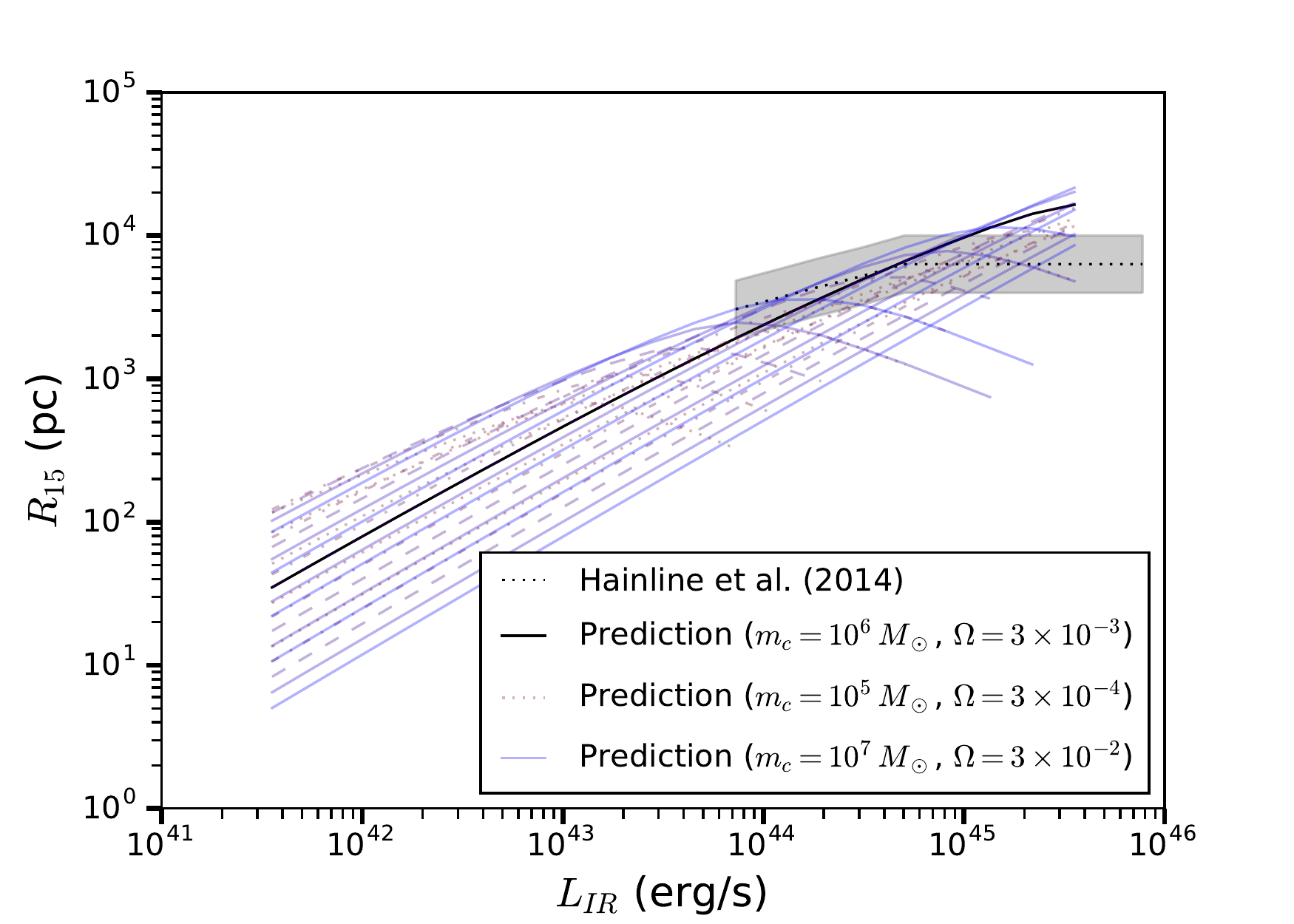}
\caption{The size-IR luminosity relationship for quasars with source luminosities in the range \SIrange{e42}{e46}{erg/s} using parameters $m_c = \SI{e6}{M_\sun}$ and $\Omega = \num{3e-3}$. The best fit line has a slope $\beta = 0.72$. Also plotted in dotted lines is the same relationship for values sampled from $\SI{e5}{M_\sun} \le m_c \le \SI{e7}{M_\sun}$ and $\num{3e-4} \le \Omega \le \num{3e-2}$. Higher $m_c$ models are shown in blue, and higher $\Omega$ models are shown in thicker lines. \label{fig:size_ir_obscuration}}
\end{figure}

This second possibility can be tested by adding a factor of $(1-\Omega_\text{tot}(r))$ to the luminosity density to account for obscuration by clouds, where $\Omega_\text{tot}(r)$ is the total covering factor of clouds up to radius $r$. This adjustment does lead to a sharp transition in the slope of the size-luminosity relation, as shown for a family of parameter values in Figure \ref{fig:size_ir_obscuration}. This suggests that an effect of this kind is responsible for the flattening observed in \cite{hainline2014}, which indicates that narrow line regions become saturated by optically thick clouds at high luminosities.

\subsection{Spatial Distribution of NLR Gas}\label{sec:simplifications}

Our model assumes the cloud density follows an inverse square law, which holds in the case of a steady-state wind. Another possibility is that of passively illuminated gas present in the galaxy before the onset of AGN activity. This would lead to a density profile which is more shallow than $n_c\propto r^{-2}$, which would have implications for all of the model results. If $n_c$ does not follow an inverse square law the covering factor $\Omega(r)$ would be non-constant, so observations of $\Omega$ may help to deduce the correct scaling.

We explore the effects of the scaling of $n_c(r)$ and $\Omega(r)$ using our model.
With the steady-state wind assumption, we find
\begin{equation}
    \d{L_{\text{\oiii}}}{r} = 4\pi r^2j_\text{\oiii} \simeq \left.\d{L_{\text{\oiii}}}{r}\right|_{r=r_s} \left(\frac{r}{r_s}\right)^{-1/2}.
\end{equation}
This means $L_\text{\oiii}$ scales with $r_\text{break}^{1/2}$, and since $r_\text{break} \propto L^{1/2}$, this represents a factor of $L^{1/4}$ in $L_\text{\oiii}$. If $n_c(r) \propto r^{-2+\delta}$, then $\Omega(r) \propto r^{\delta}$, and $\d{L_{\text{\oiii}}}{r} \propto r^{\delta-1/2}$. This adds an additional factor of $L^\delta$ to
$L_\text{\oiii}$, contributing to an overall flattening of the size-luminosity relationship. 

\begin{figure}
\includegraphics[width=\linewidth]{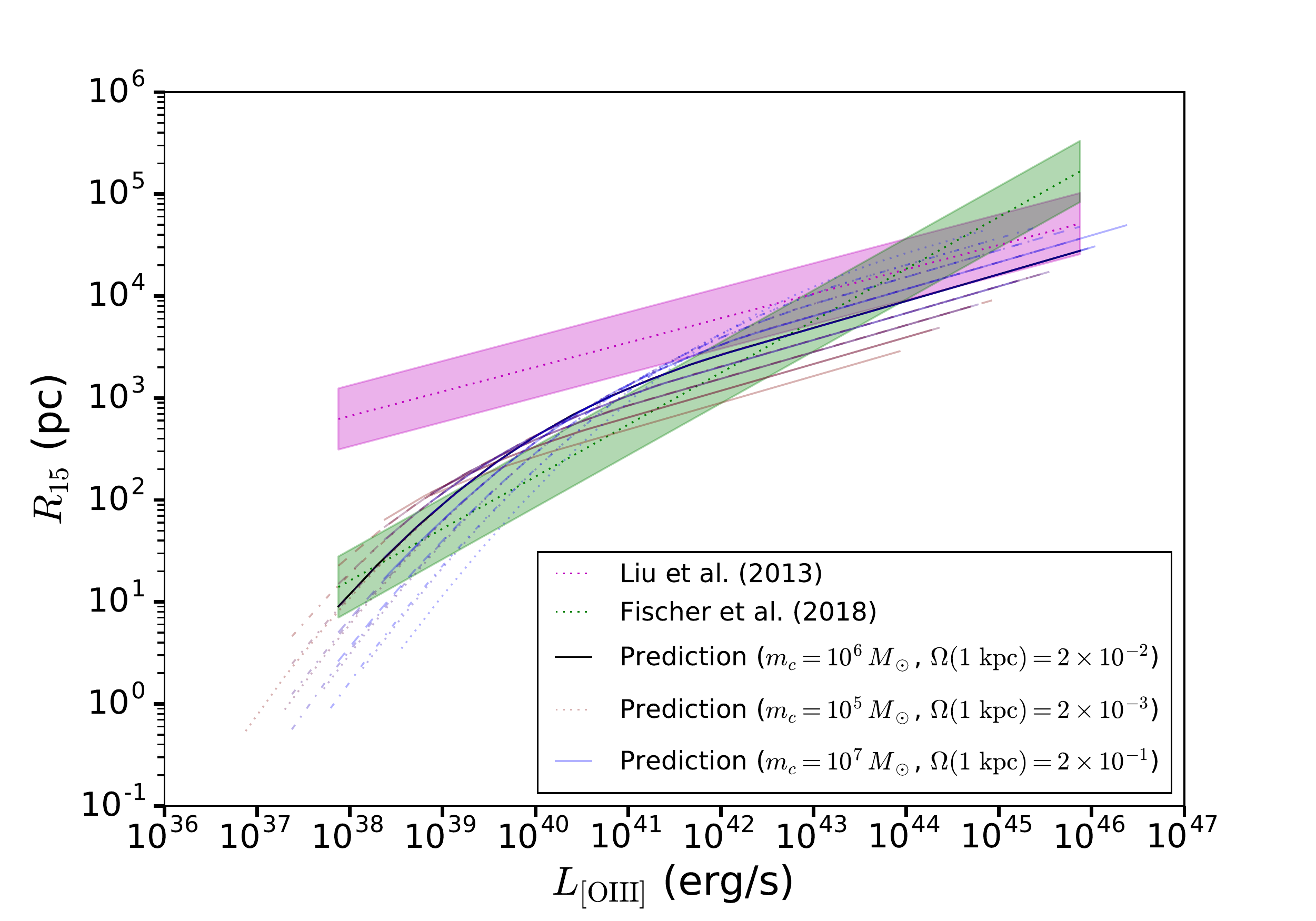}
\caption{The size-\oiii\ luminosity relationship for quasars with source luminosities in the range \SIrange{e42}{e46}{erg/s} using parameters $m_c = \SI{e6}{M_\sun}$ and $\Omega = \num{2e-2}(r/\SI{1}{kpc})$. Also plotted in dotted lines is the same relationship for values sampled from $\SI{e5}{M_\sun} \le m_c \le \SI{e7}{M_\sun}$ and $\num{2e-3} \le \Omega(\SI{1}{kpc}) \le \num{2e-1}$. Higher $m_c$ models are shown in blue, and higher $\Omega$ models are shown in thicker lines. Observed slopes in the purple and green regions are from \citet{liu2013}\replaced{, using \citet{fraquelli2003} and \citet{schmitt2003} respectively as the low-luminosity measurements.}{ and \citet{fisc18} respectively.} \label{fig:size_luminosity_flatter}}
\end{figure}

This approximate argument is confirmed by the numerical model, which also shows that a scaling of $n_c(r) \propto r^{-1}$ is necessary in a population of identical clouds to reproduce the scaling $R_{15}\propto L_\text{\oiii}^{0.25}$ observed in \cite{liu2013}. In fact, this density scaling leads to agreement between our model and the observations in both the slope and the value of $R_{15}$ at high luminosities, as shown in Figure \ref{fig:size_luminosity_flatter}.

The quantitative agreement between our models and the data above $\SI{e42}{erg/s}$ and the change in slope below this value suggests a closer look at the low-luminosity objects in \cite{liu2013}. As already discussed in Section \ref{sec:o3_results}, almost all of the objects below $\SI{e42}{erg/s}$ come from long-slit spectroscopy of low-redshift Seyfert 2 galaxies \citep{fraquelli2003}. The values of $R_{15}$ for these objects are estimated using the best-fit power laws for their \oiii\ profiles, extrapolated to low surface-brightness values. These measurements are difficult at low luminosities, particularly because of contamination from star formation, which may flatten the observed surface brightness profile and artificially increase the calculated $R_{15}$ \citep{sun2018}, explaining the difference between the observed and the model slopes of the size-luminosity relationship. In contrast, at high luminosities, where contamination effects are much less important and where the relationship is better measured, our models show quantitative agreement with the measurements. 

An additional source of ionization not taken into account in our calculations is shocks, which can be present if the NLR is participating in an outflow. Evidence for shocks is present in the line ratios of radio-quiet type 2 quasars \citep{zaka14}. Preliminary calculations suggest that in the clouds directly exposed to the photo-ionizing radiation the photo-ionization dwarfs shock ionization and therefore most of our calculations should remain unaffected. But unlike the ionizing radiation, which cannot affect the gas shadowed by the torus, winds can curve around obstacles, so shock ionization can contribute outside of the conical photo-ionized regions. 

\subsection{Constraints on Infrared Parameters}
\label{sec:ir_parameters}

Our model for infrared emission from the dusty torus requires two parameters: the covering fraction $f$ of the torus, and the fraction $\varepsilon$ of quasar light absorbed in the torus which is reradiated at infrared wavelengths. The value of $f$ is the probability of a line of sight intersecting the torus, or equivalently, the probability of an observer seeing a type 2 object. The relative frequency of type 1 and type 2 objects is controversial, with observations implying values of $f$ between 0.3 \citep{treister2008} and 0.5 \citep{reyes2008,lawr10}. There is also evidence indicating that $f$ may vary with the bolometric luminosity \citep{treister2008}. We consider the luminous end of the quasar distribution and fit a single value of $f$, which should then be interpreted as the type 2 fraction at these luminosities. 

We use observations of $L_\text{\oiii}$ and $L_{\SI{8}{\mu m}}$ (interpolated between the published values $L_{\SI{5}{\mu m}}$ and $L_{\SI{12}{\mu m}}$) in a sample of 2920 type 2 quasars \citep{yuan2016}, as well as observations of $L_{\SI{8}{\mu m}}$ and $R_{15}$ in 30 type 2 quasars \citep{hainline2014}. Our model can predict values for each of these quantities, so the $L_\text{\oiii}/L_{\SI{8}{\mu m}}$ and $R_{15}/L_{\SI{8}{\mu m}}$ relationships provide two independent constraints on the parameters $f$ and $\varepsilon$. We assume throughout our fiducial values for the NLR cloud parameters, $m_c = \SI{e6}{M_\sun}$ and $\Omega = \num{3e-3}$.

\begin{figure}
\includegraphics[width=\linewidth]{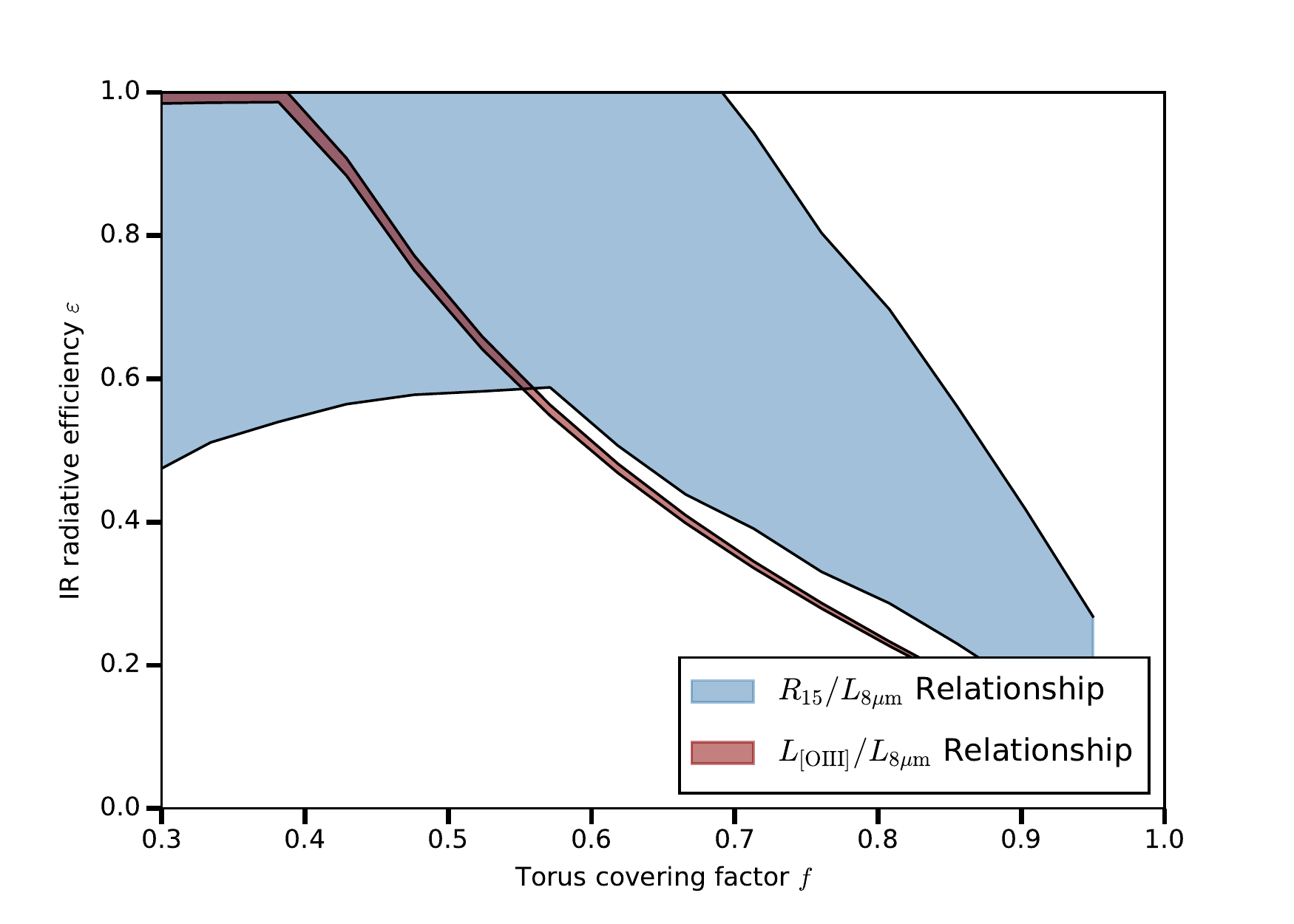}
\caption{The 1$\sigma$ confidence regions for the parameters $f$ and $\epsilon$, based on fits of our model to the $L_\text{\oiii}/L_{\SI{8}{\mu m}}$ relationship \citep{yuan2016} and the $R_{15}/L_{\SI{8}{\mu m}}$ relationship \citep{hainline2014}. Neither observation constrains the individual parameters, but the overlap of the two confidence regions suggests $0.4<f<0.55$ and $0.6<\varepsilon<1$.  \label{fig:ir_parameters}}
\end{figure}

Figure \ref{fig:ir_parameters} shows 1$\sigma$ confidence regions based on both observations. Neither observation taken alone can constrain the individual parameters, since each relationship is determined by some algebraic combination of $f$ and $\varepsilon$. However, taken together, the overlap of the confidence regions suggests $0.4<f<0.55$ and $0.6<\varepsilon<1$. For the purposes of the calculations in Section \ref{sec:ir_results}, we take $f = 0.5$, which corresponds to $\varepsilon =0.71$.

\cite{treister2008} use a somewhat similar method to estimate $f$ based on the ratio of infrared to bolometric luminosity in type 1 AGN. They estimate a parameter related to $\varepsilon$ using radiative transfer models of the torus, whereas we instead supplement the luminosity measurements with the size $R_{15}$. Since the measurements of \cite{hainline2014} are at high luminosities, our value of $f$ is best compared with the high luminosity estimate of \cite{treister2008}, $f\sim 0.3$. The difference between their value and ours may be due in part to a bias: type 2 objects are more likely to be selected from the population with a higher value of $f$ and therefore $f$ is biased high if measured from a type 2 sample, and vice versa for type 1s \citep{obied2016}. 

The extent of this bias depends on the distribution of $f$ within the population. If $p(\theta)$ is the underlying distribution of half-opening angles of ionization cones in the quasar population, then the type 2 fraction is $f=\langle\cos\theta\rangle$, where the averages denote convolution with the $p(\theta)$ probability distribution. An observer measuring this value from a type 2 population would measure $f_{\rm t2}=\langle\cos^2\theta\rangle/\langle\cos\theta\rangle$. An observer measuring this value from a type 1 population would measure $f_{\rm t1}=\langle\cos\theta(1-\cos\theta)\rangle/\langle(1-\cos\theta)\rangle$. Thus, taken at face value, our measurement $f_{\rm t2}=0.5$ and $f_{\rm t1}=0.3$ from \citet{treister2008} can be reconciled at $f\sim 0.45$ if the range of half-opening angles in the population is broad enough.

\subsection{Comparison of Mass Estimates}\label{sec:mass_discussion}

The mass of the gas involved in quasar feedback is critical for determining the role of AGN in galaxy formation. Since \oiii\ lines can be enhanced at distances much greater than the scale of mechanical feedback, we cannot determine the mass of gas involved in outflows \citep{keel2017b}. The value of our radiative model is in being able to estimate the total mass of the illuminated clouds lying within the NLR bicone \citep{stor10,das06}, even despite the fact that only a fraction of it is producing the observed emission lines. We are able to do this because our model allows us to place constraints on the masses of the illuminated clouds by using all available information on the NLR surface brightness profiles and sizes.

The multiphase nature of the illuminated gas makes it difficult to determine its mass. In particular, the neutral gas behind the ionization front is not observable in optical emission lines \citep{riff14}. Our study confirms that the matter-bounded transition is important in the explanation of some features of size-luminosity observations, including the higher slope of the size-IR luminosity relation compared to that of the size-\oiii\ luminosity relation. Additionally, the small dispersion around the size-\oiii\ luminosity relation results from most of the cloud population being ionization-bounded. Therefore, our model suggests that mass estimates based on optical emission lines miss a portion of the gas clouds.

\begin{figure}
\includegraphics[width=\linewidth]{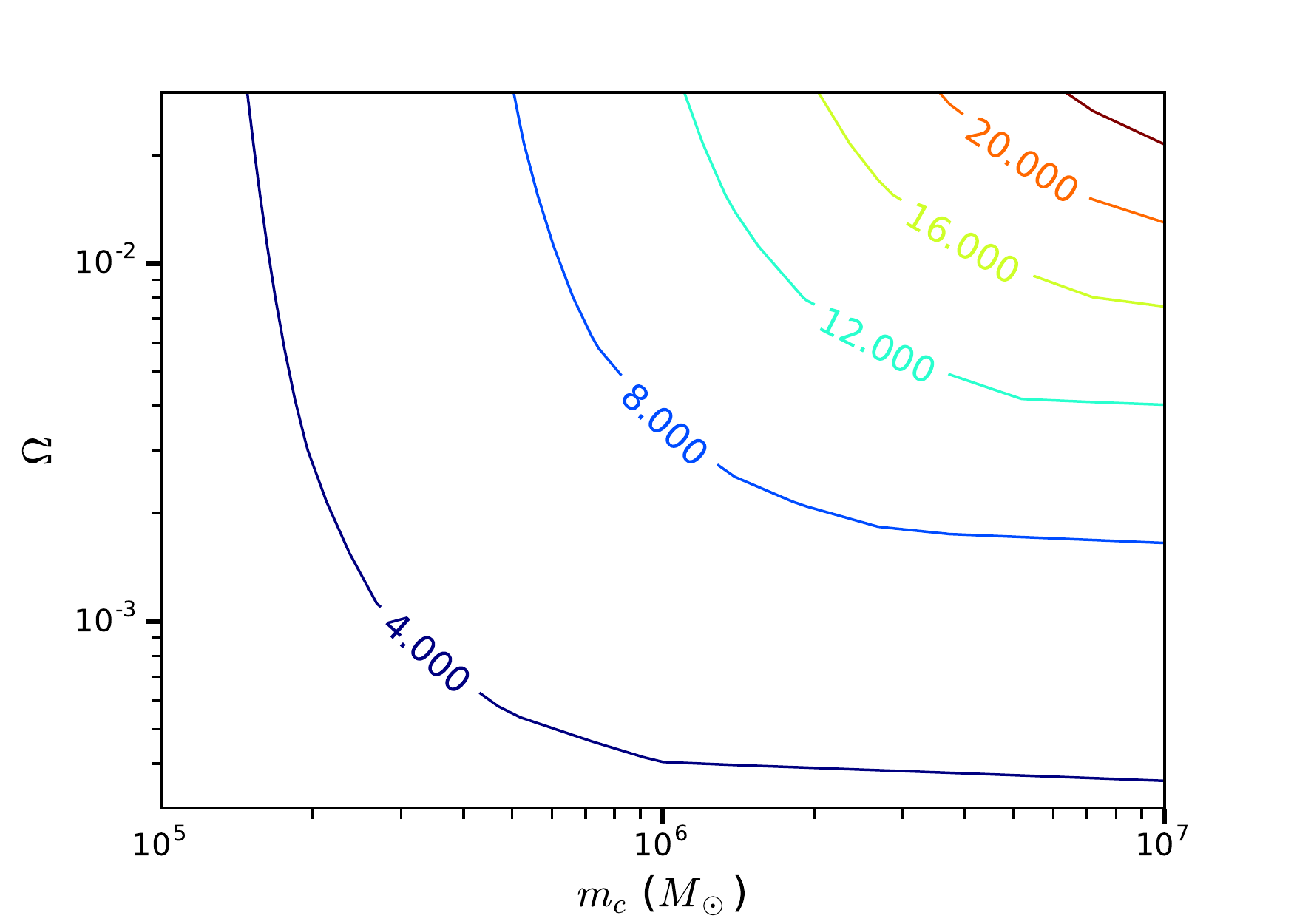}
\caption{The dependence of the $M_{15}/M_H$ ratio on cloud mass $m_c$ and covering factor $\Omega$. Parameters are shown in the ranges $\SI{e5}{M_\sun}<m_c<\SI{e7}{M_\sun}$ and $\num{3e-4}<\Omega<\num{3e-2}$. The ratio $M_{15}/M_H$ varies between 2.5 and 32.0. \label{fig:mass_comparison}}
\end{figure}

One possible way to measure the mass of the illuminated clouds is to use the hydrogen recombination lines to place a lower bound on the mass \citep{nesvadba2006}. In ionization-bounded clouds, the gas behind the ionization front will not be detected by this method. Therefore, our mass value $M_{15}$ ought to be higher than the H$\alpha$-based value $M_H$, since it includes this additional component of the gas clouds. Although in standard diagnostic diagrams (e.g., \citet{kewl06}) the cutoff \oiii/H$\beta$ ratio for AGN can be as low as 3, in luminous AGN such as those studied by \citet{liu2013} this ratio is typically 10 and above. Using the usual ratio $\text{H}\alpha/\text{H}\beta \sim 2.9$, and using the electron density $n_e \sim \SI{100}{cm^{-3}}$ which is typical at the ionization front \citep{liu2013b}, the mass value derived by \cite{nesvadba2006} can be expressed as
\begin{equation}
    \frac{M_H}{M_\sun} = \num{2.82e9}\times\frac{L_\text{\oiii}}{\SI{e44}{erg/s}}.
\end{equation}
The ratio $M_{15}/M_H$ is plotted in Figure \ref{fig:mass_comparison} for model parameters $\SI{e5}{M_\sun}<m_c<\SI{e7}{M_\sun}$ and $\num{3e-4}<\Omega<\num{3e-2}$. In this region, the ratio falls between 2.5 and 32.0, confirming that the estimate $M_H$ misses a significant fraction of the illuminated cloud mass. Moreover, the agreement of the predictions in Figure \ref{fig:size_luminosity} with observations at high luminosities suggests that the cloud masses reach at least $\SI{e6}{M_\sun}$, and at this cloud mass the ratio is 3.2 at minimum.

Additionally, the dependence of the ratio $M_{15}/M_H$ on the parameters suggests that the missing mass is the mass behind the ionization front. A model with a higher cloud mass has a greater $M_{15}/M_H$ ratio, since there is more mass behind each ionization front. Likewise, a model with a higher cloud covering factor has a greater $M_{15}/M_H$ ratio, since there are more matter-bounded clouds which are missed in optical emission lines.

The transition from ionization-bounded to matter-bounded regime is useful in determining the masses of individual clouds, which is the foundation of the mass measurement method proposed by \citet{liu2013b}. In this method, the product of cloud mass and cloud number density is constrained at the matter-bounded transition distance by the measured surface brightness at this distance. This gives the density at the matter-bounded transition distance, which can be used to compute a mass or energy flux through the sphere at this distance. The critically ionization-bounded clouds near this distance are ideal for this measurement, since they contain little to no mass behind the ionization front \citep{liu2013b}. As a result, \citet{liu2013b} derive surprisingly high mass outflow rates of 1000-2000 M$_{\odot}$ yr$^{-1}$ in ionized gas alone. Figure \ref{fig:mass_distribution} shows that a maximum mass of $m_u = \SI{6e6}{M_\odot}$ agrees well with the observed matter-bounded transition at $7.0\pm\SI{1.8}{kpc}$ \citep{liu2013}.

This method is equivalent to assuming that all NLR have the same mass and that they all become matter-bounded at the same distance. But the matter-bounded transition is observed to take place over several kiloparsecs, and the position depends on the mass as $r_\text{break}\propto m_c^{1/4}$, so the mass distribution is clearly rather broad. Furthermore, variations in the density profile $n_c(r)$ and in the mass distribution $\d{\Omega}{m}$ affect the surface brightness profile in similar ways, so it is not possible to derive the cloud mass distribution from the surface brightness profiles. These difficulties affect the mass estimate using the method by \cite{liu2013b}. 

Finally, for completeness, another method to measure the energetics of ionized gas winds is to forego the mass estimates entirely and use a Sedov-Taylor-like approach to calculate the injected power using the size and expansion rate of the wind \citep{nesvadba2006, greene2012}. This method does not suffer from many of the uncertainties associated with the unknown mass distribution of the NLR clouds, but it depends critically on the largely unknown density of ambient medium into which the wind propagates. Comparison of this method with the hydrogen-recombination method reveals that, unsurprisingly, the latter results in strict lower limits on mass outflow rates, since large amounts of mass are hidden behind the ionization front \citep{greene2012}. 

\section{Conclusion}\label{sec:conclusions}

Observations of outflows of ionized gas have become one of the major pathways to characterizing AGN feedback, and understanding the physics of the photo-ionized narrow-line regions (NLRs) is essential for this effort to succeed. In this paper we develop a physical model for the dense gas clouds in the NLR which we test against several observations of radiative feedback, including the size-luminosity relationship of the NLRs. 

Our model in its simplest formulation depends only on the central ionizing luminosity $L$, the covering factor of clouds $\Omega$ at a fiducial distance from the nucleus, and the mass of clouds $m_c$. The model of infrared luminosity, which assumes a conical geometry in the unification model, also depends on the torus covering factor $f$ and its infrared emissivity $\varepsilon$. 

In the framework of this model, we can offer explanations of several aspects of the observed size-luminosity relations. First, we find that for a wide range of model parameters there exists a tight relationship between the NLR size and various luminosity measures. The relationship owes its existence to the presence of ionization-bounded clouds. Adding extra mass behind the ionization front for such clouds does not change any of the observables, leading to the insensitivity of the model to the cloud mass distribution, as long as some high-mass clouds are present. 

Second, at high AGN luminosities the model size-luminosity relationship (which in the ionization-bounded regime has no free parameters) is in quantitative agreement with the observed ones \citep{liu2013, hainline2014}. At lower AGN luminosities our model predicts smaller NLR sizes than those that are observed. This is likely due to additional sources of extended ionization, such as star formation \citep{sun2018}. 

We find that the sizes of the NLR in the highest luminosity AGN are set by photo-ionization physics in ionization-bounded clouds (those that are optically thick to ionizing radiation) with masses $M\ga 10^6$ M$_{\odot}$. But the transition of NLR clouds from the ionization-bounded to the matter-bounded regime plays an important role in several observables. In particular, the steep surface brightness decline of the \oiii\ nebulae is naturally reproduced by clouds of a range of masses which transition to the matter-bounded regime at different distances. Therefore, these surface brightness profiles offer a potential probe of the mass distribution of NLR clouds. A Milky-Way-like cloud mass distribution produces a reasonable agreement with the observed \oiii\ profiles.  

While our model does not well reproduce the slope of the size-luminosity relationship, we are able to reproduce the observation of \cite{hainline2014} that the size-IR luminosity relation is steeper than the size-\oiii\ luminosity relation. This is a result of the matter-bounded transition in NLR clouds. The distance at which the matter-bounded transition takes place increases with the source luminosity, so brighter AGN have more clouds available for \oiii\ emission. This leads to a steeper scaling of $L_{\text{\oiii}}$ with $L$ in comparison with the direct proportionality of $L_\text{IR}$ to $L$, and this corresponds to the size-IR slope exceeding the size-\oiii\ slope. The flattening of the size-IR luminosity relation observed in \cite{hainline2014}, where sizes do not increase with luminosity beyond $R\sim 10$ kpc, may be due to the saturation of the quasar sky by optically thick clouds, preventing escape of ionizing radiation. 

Our model also provides a new method of determining the typical mass of the NLR. Mass estimates that rely on individual surface brightness measurements of optical emission lines  are insensitive to the significant quantity of gas which lies behind the ionization fronts of clouds. Our model includes this portion of the clouds, and the mass distribution of clouds necessary for this correction can be constrained to some extent by comparing the model results with observations of the matter-bounded transition distance and \oiii\ surface brightness profiles. With current constraints, we find that the NLR masses are 3--30 times higher than those estimated from hydrogen recombination lines. 

As a byproduct of simultaneously modeling the size-luminosity relationship and the \oiii-IR correlation for type 2 quasars, we are able to determine the obscuration covering factor (equal to the fraction of type 2 AGN in the population) to be $f=0.5$. Correcting this value for a mild type 2 selection bias brings it down to $f=0.45$. This measurement provides independent support for several lines of observation evidence suggesting that half of the AGN population is obscured even on the high-luminosity end.  

\acknowledgments

The authors are grateful to J. Stern and A.-L. Sun for many useful discussions. Support for this work was provided in part by the National Aeronautics and Space Administration through Chandra Award Number AR7-18011X issued by the Chandra X-ray Observatory Center, which is operated by the Smithsonian Astrophysical Observatory for and on behalf of the National Aeronautics Space Administration under contract NAS8-03060. NLZ acknowledges support by the Catalyst award of the Johns Hopkins University.




\bibliographystyle{aasjournal}



\end{document}